\documentclass{aastex}

\slugcomment{Submitted to ApJ Lett.}
\def\etal{{\it et al.~}}

\shorttitle{Radio Variability of Sagittarius A*}
\shortauthors{Zhao et al.}

\begin{document}


\title{Radio Variability of Sagittarius A*
- A 106 Day Cycle}


\author{Jun-Hui Zhao}
\affil{Harvard-Smithsonian CfA, 60 Garden St, MS 78, Cambridge, MA 02138}
\email{jzhao@cfa.harvard.edu}

\and
\author{ G. C. Bower \& W. M. Goss}
\affil{NRAO-AOC, P. O. Box 0, Socorro, NM 87801}
\email{gbower@nrao.edu \& mgoss@nrao.edu }



\begin{abstract}

We report the presence of a 106-day cycle in the radio 
variability of Sgr A*  based on an analysis of
data observed with 
the Very Large Array (VLA) over the past 20 years.
The pulsed signal is most clearly seen at 1.3 cm with a
ratio of cycle frequency to frequency width 
$f/\Delta f= 2.2\pm0.3$.  The periodic signal is also clearly 
observed at 2 cm. 
At 3.6 cm
the detection of a periodic signal is marginal.
No significant periodicity is detected at both
6 and 20 cm.  Since the sampling function is irregular we performed a number
of tests to insure that the observed periodicity is not the result
of noise.
Similar results were found for a maximum entropy
method and periodogram with CLEAN method.  The probability
of false detection for several different noise distributions
is less than 5\%
based on Monte Carlo tests.  
The radio properties of the pulsed component
at 1.3 cm
are spectral index $\alpha\sim1.0\pm 0.1$ (for $S\propto\nu^\alpha$),
amplitude $\Delta S=0.42\pm 0.04 {\rm\ Jy}$
and characteristic time scale $\Delta t_{FWHM}\approx 25\pm5$ days.
The lack of VLBI detection of a secondary component suggests that
the variability occurs within 
Sgr A* on a scale of $\sim5$ AU, suggesting
an instability of the accretion disk.

\end{abstract}




\section{Introduction}
The compact radio source 
Sagittarius A* is suggested to be associated with a supermassive
black hole  at the Galactic Center (Eckart \& Genzel \etal 1997, Ghez \etal 1998,  Backer \& 
Sramek 1999, Reid \etal 1999).  
The flux density variability of Sgr A* 
has been puzzling since the discovery of this intriguing radio compact 
source at the center of the Galaxy in 1974 (Brown \& Lo 1982; Zhao \etal 1989). 
We monitored Sgr A* with the 
VLA during the period of 1990 to 1993.  
Fluctuations in flux density 
suggested that the amplitude of variation increased towards 
short wavelengths and that the rate of outbursts
is about three per year
(Zhao \etal 1992; Zhao \& Goss 1993).
Large amplitude fluctuations in the flux density 
have been observed 
at millimeter wavelengths (Backer \& Wright 
1993; Tsuboi \etal 1999).
Based on the radio monitoring data obtained with
the 3.5 km Green Bank
Interferometer at 11 and 3.6 cm,
Falcke (1999) reported that at both wavelengths a characteristic time scale
of 50 to 200 days is observed while the structure function of 11 cm data
suggests a quasi-periodic variation with a period of 57 
days.

\section{Observations \& Data Reduction}

Regular observations were made during 1990.1-1991.5
with various sampling intervals in a range from 1 day to 28 days
at 3.6, 2 and 1.3 cm in the A, B, C, D and the hybrid configurations
of the VLA.
Less regular observations
at the
same frequencies were made at the same frequencies between
1991.5 and 1993.5.  The largest gap between observations
in these data was 120 days.  
In addition to these regularly sampled data from 1990 to 1993 discussed
previously by Zhao \etal 1992,
we have collected all the observations of Sgr A* at 20, 6, 3.6, 2 and 1.3 cm
from the VLA archive for the past two decades.  
We accepted only the 
observations  with baselines long enough
($>80 k\lambda$) to separate Sgr A* 
from the extended HII emission. 
This limit corresponds to $3^{\prime\prime}$ resolution.

Primary flux density calibration 
was performed in the standard way using 3C 48 or 3C 286.
The flux density scale was then bootstrapped
to one of several nearby compact radio sources, typically J1733-130,
J1744-312 or J1751-253.   
Most of the flux densities presented in this paper were
determined in the visibility domain, from the long baseline data.
In a few cases in which the length of baselines
was marginal in separating the  point
source in the visibility domain we determined the flux densities of Sgr A* 
in the image plane.
Then, the flux density of Sgr A*
was measured  by subtracting the 
confusion from the extended emission.  In the cases in which  the telecope 
pointing was offset from Sgr A*,  
both the primary beam and synthesis
beam effects are corrected.

Figure~1  shows the radio light curves at the five  wavelengths
for Sgr A*.
At all wavelengths determination of the absolute flux density scale
dominates the errors.
There are also contributions due to the thermal noise,
separation from the extended HII region Sgr A West for low angular resolution data
and uncertainty in the atmospheric opacity at 2 and 1.3 cm.
Uncertainty at long wavelengths is  less than 5\%.
For long tracks, this uncertainty can be minimized
by removing the attenuation of the gain as a function of elevation.
The correction for
this effect is no more than 10\% at 1.3 cm based on several
observations with 8 hr tracking time.  Most of the observations used 
in this paper were in a snapshot mode (10-20 minutes in integration). 
For these short observations, we can assess
the uncertainty based on observations of a stable, nearby 
(4$^o$ away from Sgr A*)
calibrator J1751-253 
(Figure 1b).
The 7\% fluctuation of the flux density 
at 1.3 cm for J1751-253 
reflects the uncertainty in the calibration. 

\section{Data Analysis \& Results}

\subsection{Power Spectrum and A Period of 106 Day}

We carried out a power spectral density (PSD) analysis of the
light curves in order to identify periodic signals.  With a simple Fourier 
analysis,
the irregular sampling of the VLA archive data produces 
strong side-lobes,  
leading to confusion in the identification of spectral features. 
In addition large gaps in the sampling can produce an 
alias of a periodic signal
to lower frequency lags, weakening the power of
the true signal and producing aliased signals.
To minimize the side-lobes,  we estimated the PSD
with the maximum entropy method (MEM)
(Press \etal 1989) and with a classic periodogram augmented with CLEAN.
The results from both approaches are completely consistent.

In order to overcome the aliasing problem due to large gaps in time,
we used the following procedure.
This procedure determines the period  from well-sampled but small
data set, folds an
irregularly-sampled but larger data set with that period and then
searches for a new period.
We considered three subsets of the full light curves:
1990-91, 1990-93 and 1977-1999 ({\it i.e.}, the full set).  
We removed a third-order polynomial
from each of these data sets.  This baseline represents 
slow variations in the flux density of Sgr A*.
In the 1990-91 subset, the maximum sampling
gap (28 days) corresponds to the Nyquist critical frequency of
$f_c=2.1\times10^{-7}$ Hz. 
Using the zero-leveled subset 
1990-91, we calculated a PSD profile.  At 1.3 cm,  we found a 
spectral peak at the frequency $f=1.07\times10^{-7}$ Hz, corresponding to
a period $P=107$ days.  This frequency is well within the
range of Nyquist critical frequency.

To verify the periodic signals, we examined the 1990-93 and 1977-99 subsets
folded into $N_{cyc}$ cycles of period $P_{\rm in}$.
The 1990-93 consists of 59 observations during the period 1990.1-1993.8, 
containing $\sim1/2$ of the total  data points in 1977-99.
The maximum gaps in the sampling
are 120 and 1350 days for the subsets 1990-93 and 1977-99, respectively.  
The number $N_{cyc}$ is chosen as the largest number such
that the maximum sampling gap ($\Delta t_{max}$) in the new
folded time series is smaller than half the value of the period. 
Taking the  new folded  time series of these subsets,
we calculate PSD profiles.
Table 1 and Figure 2 summarize the results.  
The peak frequencies ($f$) derived from
these three data sets are consistent, with a mean value of
1.09$\times10^{-7}$ Hz.
The uncertainty $\sigma_f$ $\sim 3\times 10^{-9}$ Hz (3 days) 
of the peak frequency 
is estimated from the maximum deviation of the peak frequencies
derived from the three data subsets.
The FWHM width 
($\Delta f$)
of the PSD feature is 9$\times10^{-8}$ Hz
derived from both 1990-91 and 1990-93 subsets. For the 1977-99 subset,
the FWHM width is reduced by a factor of 2 
($\Delta f=5\times10^{-8}$ Hz). The uncertainty of the FWHM width due to
noise and the irregular sampling is  $\sim$15\%. 
The ratio of $f$/$\Delta f$
is $\sim2.2 \pm 0.3$.

For the 2 cm data, we followed the same procedure.
The mean peak frequency
is consistent with the result obtained from 1.3 cm.  
The power spectral feature
derived from all the data at 2 cm appears to be broadened by a factor
of $\sim$2 as compared to that of 1.3 cm. The ratio of $f$/$\Delta f$
is $\sim1.0\pm0.2$ at 2 cm.

At 3.6, 6 and 20 cm we folded the data with a period of 106 days.
We used two subsets of the data for the 3.6 cm data, 1990-1993
and 1988-1999, and the whole data sets for the 6 and 20 cm data.
At 3.6 cm a spectral feature with a period of 106 days is
marginally detected.  This feature has a ratio of peak
frequency to width $f/\Delta f  < 0.4$.
No significant features are detected in the 6 and 20 cm data.  

Using the procedure discussed above, we also folded the 1.3 cm data
into a number of cycles with a small period to check if 
the PSD feature observed is an aliased signal from
a higher frequency periodic variation. 
We have checked the periods ranging between 1 to 56 days
which is outside the highest Nyquist frequency 
$f_c=2.1\times10^{-7}$ Hz provided by the VLA observations.
No spectral features from the folded data sets
are stronger and narrower than the $1\times10^{-7}$ Hz feature.
No periodic 
signals from the calibrator J1751-253 is found confirming
 that  the periodic 
variation is not caused by calibration errors.

There is a probability that the periodic signal discussed 
above could be a false detection due to  
a combination of a random process and the 
irregular sampling function.
To provide a quantitative estimate, we have carried out 
Monte Carlo tests with data sets created from
various noise sources using
the sampling function of the real data and the procedure outlined
above.  Our noise functions included white noise, Gaussian noise
around a mean and a Poisson distribution of flares.
These flares were modeled as $\delta$ function rises in flux density
that occurred with a probability $p$ and decayed with an exponential time
constant $t_d$.  We searched the ranges $0.001 < p < 0.27$ and
$1 < t_d < 90 $ days. 
We also considered sets of data in which we reordered (or 
scrambled) the actual measured flux densities.
This test assumes no knowledge of the parent distribution of the flux
density.
Based on the 1990-93 data at 1.3cm, we find
that the probability of false detection due to a random
process is less than 5\% for all of these cases.  False detection
occurs when the ratio $f/ \Delta f$ of the noise data set exceeds
that of the true data set with the same sampling function.

\subsection{The Mean Profiles of the 106 Day Cycle}

Figure 3 shows  the mean profiles of the 106 day cycle
constructed by
folding the time series data (1990.1-1993.8) into a 106 day period.
Again, the long term baselines were removed before  folding.
The zero levels reflect the mean flux densities of Sgr A*
as listed in the caption to Figure 1. 
The 0-phase in the plots corresponds to a reference day
(4 Dec 1991). Because of the uncertainty in period (3\% from the 1.3 cm data),
the phase uncertainty for folding long time baseline data becomes large.
The maximum uncertainty of the phase  in the mean profile for the period
1990.1-1993.8 is $\sim$20 days.
The radio properties of the 106 day
cycle are summarized in Table 1.
Both the absolute ($\Delta S$) and fractional 
 ($\Delta S/ S$) amplitudes  of the ``pulsed''
 component appear to increase
towards short wavelengths. The mean spectral index in the ``pulsed'' component
 is $\alpha~\approx 1.0\pm 0.1$ (S$\propto\nu^\alpha$). 
The phase transition from minimum
to maximum appears to be sharper at 1.3 cm than 2 cm, although
we could not observe any significant phase offsets.
The FWHM ($\Delta t \pm \sigma_{\Delta t}$) of the mean profile is roughly 
$25\pm5$  and $40\pm10$ days
for 1.3 and 2 cm, respectively. There are multiple peaks in the 3.6 profile
indicating the typical time scale ($\Delta t$) of individual periodic events  is comparable
to the the period (106 days) at this wavelength.
In addition, the FWHM width of the periodic feature increases as
the wavelength increases as shown in Figure 2. The periodic 
fluctuation appears to  diminish at the longer
wavelengths.

\section{Discussion}
If the 106 day cycle is related to an orbital emitting object
around the massive black hole, its distance to the central
mass would be 1200$\times$ R$_g$ (60 AU), where
R$_g = 7.5\times 10^{11}$ cm for 2.5$\times10^6$ M$_\odot$. 
At a distance of 8 kpc, 60 AU corresponds
to 8 mas.  A 200 mJy compact object separated by 8 mas
from Sgr A* is easily detected with the VLBA at wavelengths
shorter than 3.6 cm. 
There is no evidence for a companion source in images
of Sgr A* at any wavelength (e.g., Bower \& Backer 1998).
We can also demonstrate
 the unlikelihood of the specific case of an eccentric binary pair
in which a wind from the secondary eclipses the primary as
in the case for the pulsar
PSR 1259-63 (Johnson \etal 1992).
The opacity of free-free absorption $\tau_{f-f} \sim \lambda^{2.1}$
suggests that  
the periodic variability due to an orbiting body would be enhanced at 
longer wavelengths.
At $\lambda \ge$ 6 cm, Sgr A* would be completely absorbed
in the eclipsed phase if $\tau\sim 0.5$ as inferred from 
the fraction of 40\% in the flux density variation at 1.3 cm. 
Thus, the periodic variability is more likely intrinsic to Sgr A* 
and probably occurs within a few hundred R$_g$ of 
the massive black hole.  


 Quasiperiodic variability
in radio flux density can be produced through
jets or collimated flows.
With a wealth of data observed at wavelengths between radio to X-ray,
the microquasar GRS 1915+105 provides an excellent case 
showing two distinct states ( ``plateau'' and ``flare'')
of the accretion process in  
a stellar mass black hole,
 (Mirabel \& Rodriguez 1999;
 Dhawan \etal 2000). 
 The current state of Sgr A*
 is similar to the ``plateau'' state of GRS 1915+105
 in terms of flat radio spectrum, compact source size (AU scale)
 and periodic variability
 in radio flux density. This state can be  
contrasted to  the ``flare'' state  which is characterized by
 optically thin ejecta feeding large scale jets (500 AU).
 The radio variability as correlated with the soft X-ray cycle
 in the ``plateau'' state of GRS 1915+105 suggests that the radio
 oscillations  
 correspond to  the thermal-viscous instability of the accretion disk
 (Dhawan \etal 2000).
 The analog of the radio fluctuations in the two sources suggests that 
 the periodic flux density variations in Sgr A*
 are also related to an instability of the accretion disk.
 However, the X-ray luminosity ($\sim2\times10^6$ L$_\odot$)
 of GRS 1915+105 is 20 times greater than
 the Eddington limit for  a 3 M$_\odot$ black hole (Mirabel \& Rodriguez 1999)
 and the radio jets are produced by the overwhelming radiation pressure.
 On the other hand, the low X-ray luminosity ($<$ 100 L$_\odot$,
 $\sim$10 orders in magnitude below the Eddington limit
 for a 2.5$\times10^6$ M$_\odot$ object)
 of Sgr A* 
 indicates that the gravity   far exceeds the radiation pressure.
 Because of the strong gravity and weak radiation pressure, 
 the consequence
 of the gas dynamics in Sgr A* on the AU scales 
 would be different from  GRS 1915+105.
In fact, the time variability and 
the limit  on the intrinsic source size  ($<$0.5 mas or 100 R$_g$ or 5 AU) of Sgr A*
from the 7 mm observations (Lo \etal 1998 and Bower and Backer 1998)
suggest that 
any variability in the jet occurs in a region where the
gravitational field of the black hole dominates.  
Any collimations of jets or outflows 
related to the observed radio variability
appear to be  disrupted
within Sgr A* on a scale of $\sim$5 AU.
A model consisting of  a jet nozzle ({\it e.g.} Falcke 1996) seems useful to study if a 
self-consistent dynamic theory for the disk instability can be 
constructed.

A convection process is now considered in the advection 
dominated accretion flow
(ADAF)  that could well
provide a reasonable dynamic model for the accretion disk and
possible outflows (Narayan \etal
2000; Quataert \& Gruzinov 2000; Igumenshchev \& Abramowicz, 1999; 
Stone \etal, 1999). 
In this theory, hot dense bubbles are produced in the inner part of
a low-viscosity disk through convection caused by thermal instability.
The most attractive result
from the convective-ADAF theory is that quasiperiodic production of
convective bubbles  has been observed in numerical simulations,
although the observed period and the small cycle frequency width
have  not been  predicted in detail from the theory
(Igumenshchev \& Abramowicz, 1999).
The current results appear to favor the convective-ADAF model although
we can not rule out the possibilities of 
an orbiting companion object that triggers periodic flares from a jet nozzle.

\acknowledgments

We thank Ramesh Narayan for stimulating discussion concerning the convective
ADAF model, and Ron Ekers for his  helpful discussion
and his initial suggestion of the VLA monitoring program in 1980. 
We thank Don Backer for helpful discussions and 
Heino Falcke for sharing recent data from the GBI.
The Very Large Array (VLA) is operated by the National Radio Astronomy 
Observatory
(NRAO). The NRAO is a facility of the National Science Foundation operated 
under
cooperative
 agreement by Associated Universities, Inc.





\clearpage



\figcaption[fig1.ps]{a. (left panel)
The radio light curves of Sgr A* observed
with the VLA at 20, 6, 3.6, 2, 1.3 cm over the past two decades.
The solid lines connect two successive data points (dots). 
The vertical error bar indicates 1 $\sigma$. The dotted curves
indicate the third-order polynomial baselines which are removed
from the data prior to the periodic analysis as discussed
in Table 1 and  Figures 2 and 3.~~
b. (right panel) The radio light curves of Sgr A* (filled dots) and
a calibrator J1751-253 (open circles) observed during
1990-1993. The calibrator J1751-253 is a compact (less than 0.1 arcsec),
 steep spectrum source.
The flux density of J1751-253 was stable in this period as compared 
with those of Sgr A*.
The mean flux density and  standard deviation (${\rm S \pm \delta S}$) 
are 1.20$\pm$0.01, 0.470$\pm$0.009, 0.278$\pm$0.005, 0.157$\pm$0.007, 
0.116$\pm$0.008 Jy for J1751-253 and 
0.513$\pm$0.049, 0.710$\pm$0.072, 0.783$\pm$0.099, 0.99$\pm$0.18, 
1.10$\pm$0.23 Jy at 20, 6, 3.6, 2, 1.3 cm for Sgr A* during this period.
\label{fig1}}
\figcaption[fig2.ps]{The  power spectral density profiles
derived from all the data folded into 6 cycles
of the period 106 days at each wavelength. 
Each profile along the vertical axis is 
normalized by its peak value and is modified by adding
 0, 0.5, 0.75 , 1.5 and 2.0  
at 1.3, 2, 3.6, 6 and  20 cm, respectively.
A peak  near 1$\times10^{-7}$ Hz is clearly detected at 1.3,
2 cm while the detection at 3.6 cm is marginal ($<$ 3$\sigma$).
The width of the spectral feature increases
as the wavelength increases. No significant
periodicities from the 20 and 6 cm data were detected.
\label{fig2}}

\figcaption[fig3.ps]{
The mean profiles of the 106 day
cycle at 3.6, 2 and 1.3 cm are obtained by
folding the data obtained during 1990.1-1993.8 (Figure 1b)
into one single cycle. The profiles are smoothed.
The mean spectral index of the ``pulsed'' components is about
$\alpha\sim1$ ($\Delta$S$\propto\nu^\alpha$). The long-term baseline
(the third-order polynomial)
has been taken out in the PSD fitting and cycle folding
processes. The zero levels here reflect the mean flux densities of Sgr
A* as listed in the captions to Figure 1b.
\label{fig3}}





\clearpage

\begin{deluxetable}{llccccccccc}
\tabletypesize{\scriptsize}
\tablenum{1}
\tablewidth{0pt}
\tablecaption{The Results of Power Spectral Fitting for Sgr A$^*$}
\tablehead{ 
\colhead{$\lambda$ }&
                     &
\colhead{$f ~(\sigma_{f})$} &
\colhead{$\Delta f$ ($\sigma_{\Delta f}$)} &
\colhead{$f/ \Delta f$($\sigma_{f/\Delta f}$)} &
\colhead{$P~(\sigma_{P})$}&
\colhead{$\Delta S$($\sigma_{\Delta S}$)}&
\colhead{$\Delta S/ S$($\sigma_{\Delta S/S}$)} &
\colhead{$\Delta t$ ($\sigma_{\Delta t}$)}\\
(cm)&
    &
($10^{-7}$Hz) &
($10^{-7}$Hz) &
                &
(day) &
(Jy) &
(\%) &
(day) \\
}
\startdata

1.3  &&1.09 (0.03)& 0.50 (0.07) &2.2 (0.3) &  106 (3) & 0.42 (0.04) & 34 (9) 
& 25 (5)\\
 2.0 &&1.12 (0.10) & 0.80 (0.12) &1.0 (0.2) & 104 (9) &  0.28 (0.02) & 25 (6) & 
40 (10)\\
 3.6 &&1.10 (0.01)& $>$3       &$<$0.4  & 106 (1) &  0.16 (0.01) & 20 (4) & 
 $\sim$100\\


\enddata
\end{deluxetable}



\begin{thebibliography}{}
\bibitem[Backer and  Sramek (1999)]{bac78} 
Backer, D. C. and Sramek, R. A. 1999, \apj, 524, 805

\bibitem[Backer and Wright (1993)]{bak1993} Backer, D. C. and Wright, M.  1993, \apj, 417, 560

\bibitem[Bower and Backer (1998)]{bak1998} Bower,  G. C. and  Backer, D. C. 1998, \apj, 496, L97
 
\bibitem[Brown and Lo (1982)]{bro1982} Brown, R. L. and  Lo, K. Y.1982, \apj, 253, 108

\bibitem[Dhawan et al (2000)]{dha2000} Dhawan, V., Mirabel, I. F. and Rodriguez, L. F.
2000, \apj, 543, 1 nov 2000 issue

\bibitem[Eckart and Genzel (1997)]{eck1997}  Eckart, A., and Genzel, R. 1997, \mnras, 284, 576

\bibitem[Falcke (1996)]{fal1996} Falcke, H.  1996 in IAU Symp. 169, p163

\bibitem[Falcke (1999)]{fal1999} Falcke, H. 1999 in ASP Conf. Series 186, p113

\bibitem[Ghez et al (1998)]{ghe1998} Ghez, A. M., Klein, B. L., Morris, M., and Becklin, 
E. E. 1998, \apj, 509, 678

\bibitem[Igumenshchev and Abramowicz (1998)]{igu1998} 
Igumenshchev, I. V. and  Abramowicz, M. A. 1999, 
\mnras, 303, 309

\bibitem[Johnson etal (1992)]{joh1992} Johnston, S., Manchester, R. N., 
Lyne, A. G., Bailies, M., Kaspi, V. M., Qiao, G., and D'Amico, N. 1992,
\apj, 387, L37


\bibitem[Lo et al (1998)]{lok1998} Lo, K. Y., Shen, Z.,  Zhao, J.-H.and  Ho, 
P. 1998, \apjl, 508, L61

\bibitem[Mirabel and Rodriguez (1999)]{mir1999} 
Mirabel, I. F. \& Rodriguez, L. F. 1999,
\araa, 37, 409

\bibitem[Narayan et al (2000)]{nar2000} Narayan, R., Igumenshchev, I. V. and
Abramowicz, M. A. 2000, \apj, 539, 798


\bibitem[Press et al (1988)]{pre1989} Press, W. H., Plannery, B. P., 
Teukolsky, S. A. and
Vetterling, W. T. 1988, Numerical Recipes, Cambridge University Press

\bibitem[Quataert and Gruzinov (2000)]{qua2000} Quataert, E. and
 Gruzinov, A. 2000,  \apj, 539, 809

\bibitem[Reid et al (1999)]{rei1999}
Reid, M. J., Readhead, A. C. S., Vermeulen, R.C., \& Treuhaft, R. N. 1999, 
\apj, 524, 816

\bibitem[Stone et al (1999)]{sto1999}
Stone, J. M., Pringle, J. M., \& Begelmen, M. C. 1999, \mnras, 310, 1002

\bibitem[Tsuboi et al (1999)]{tsu1999} Tsuboi, M., Miyazaki, A. 
\& Tsutsumi, T. 1999, ASP Conf. Series 186, p105

\bibitem[Zhao et al (1989)]{zha1989} Zhao, J.-H., Ekers, R. D.,  Goss, W. M., Lo, K. Y.
and 
Narayan R.
1989, IAU
Symp. 136, 535.

\bibitem[Zhao \etal (1992)]{zha1992} Zhao, J.-H., Goss, W. M., Lo,  K. Y. and  Ekers, R. D. 1992, 
ASP
Conf. Series
31, 295

\bibitem[Zhao and Goss (1993)]{zha1993} Zhao, J.-H. and Goss, W. M. 1993
Sub-arcsecond Radio Astronomy, R.J. Davis and R. S. Booth,
 Cambridge University
  Press, 38



\end{thebibliography}
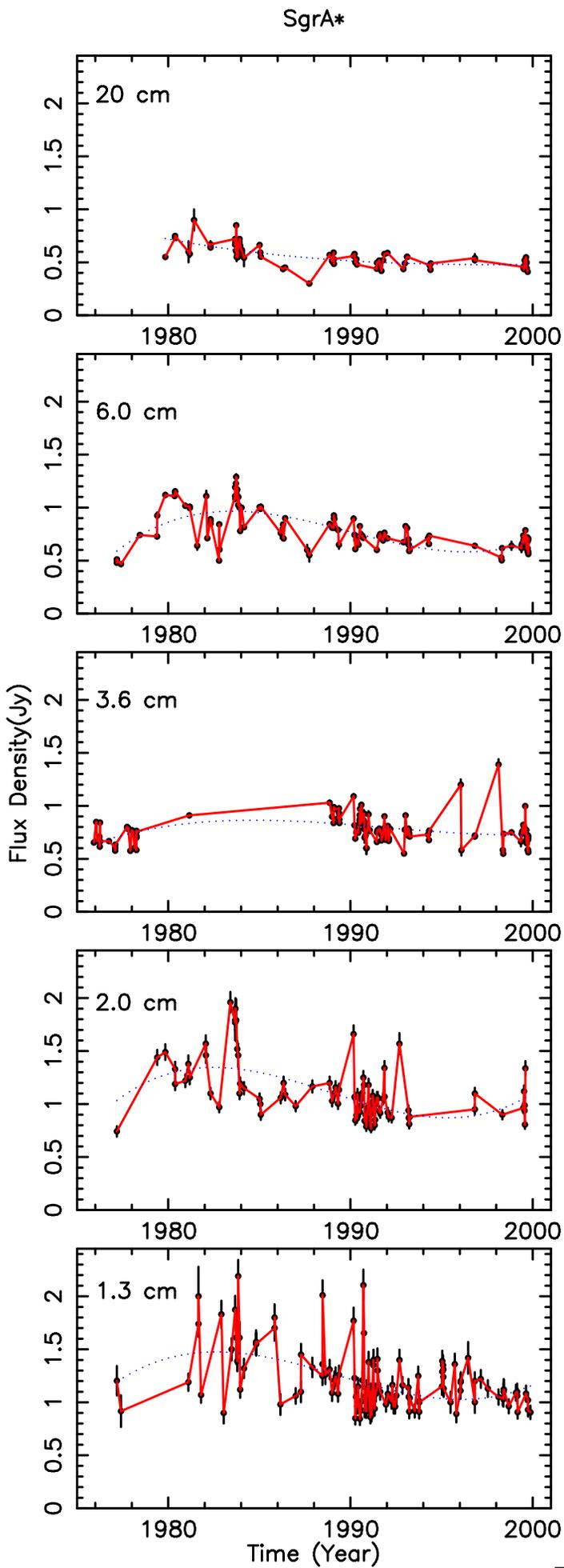
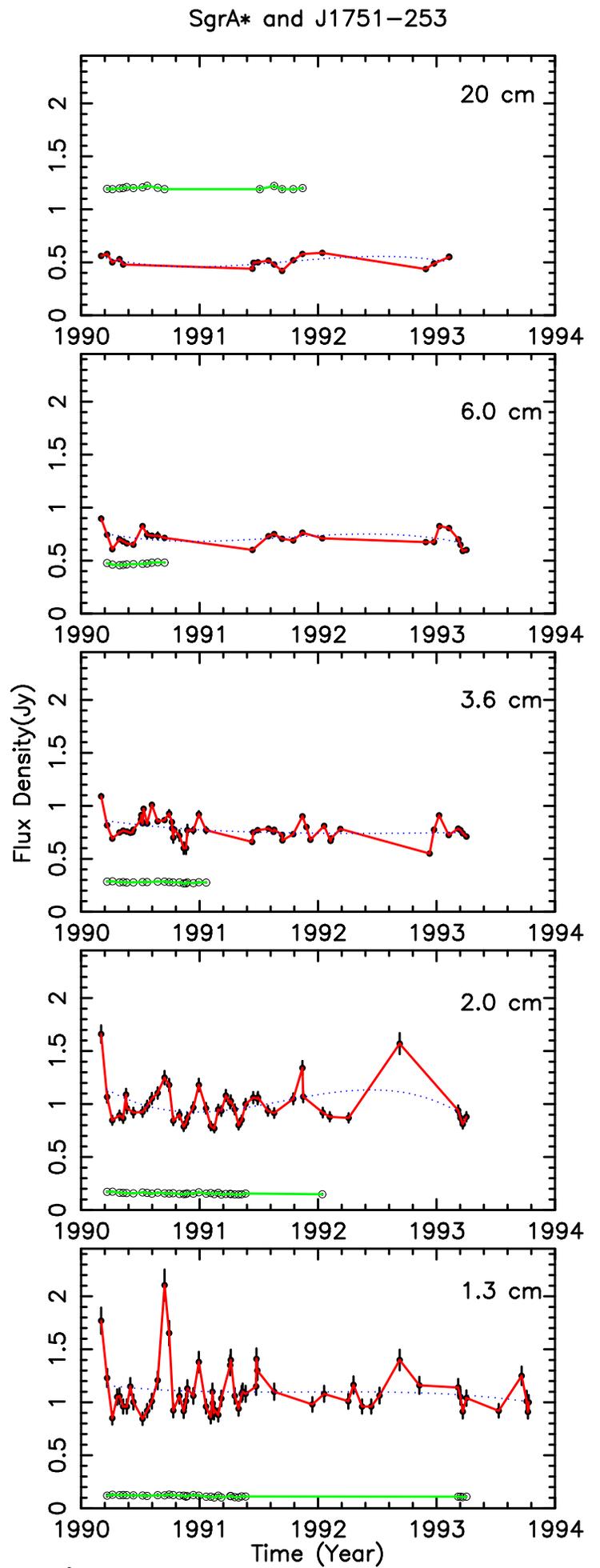
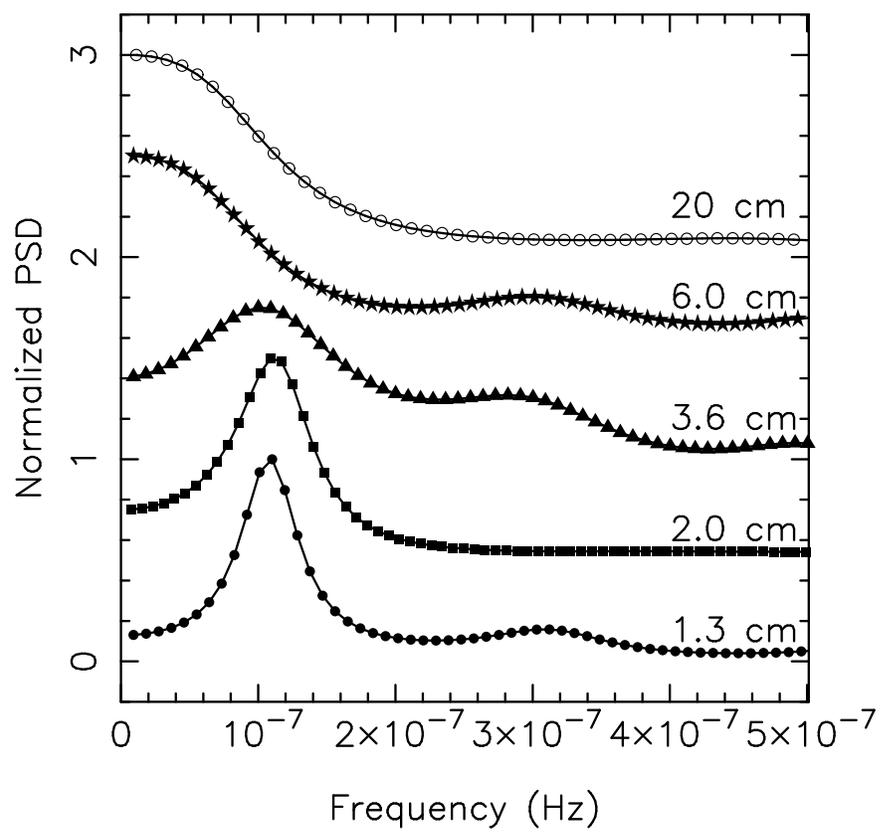
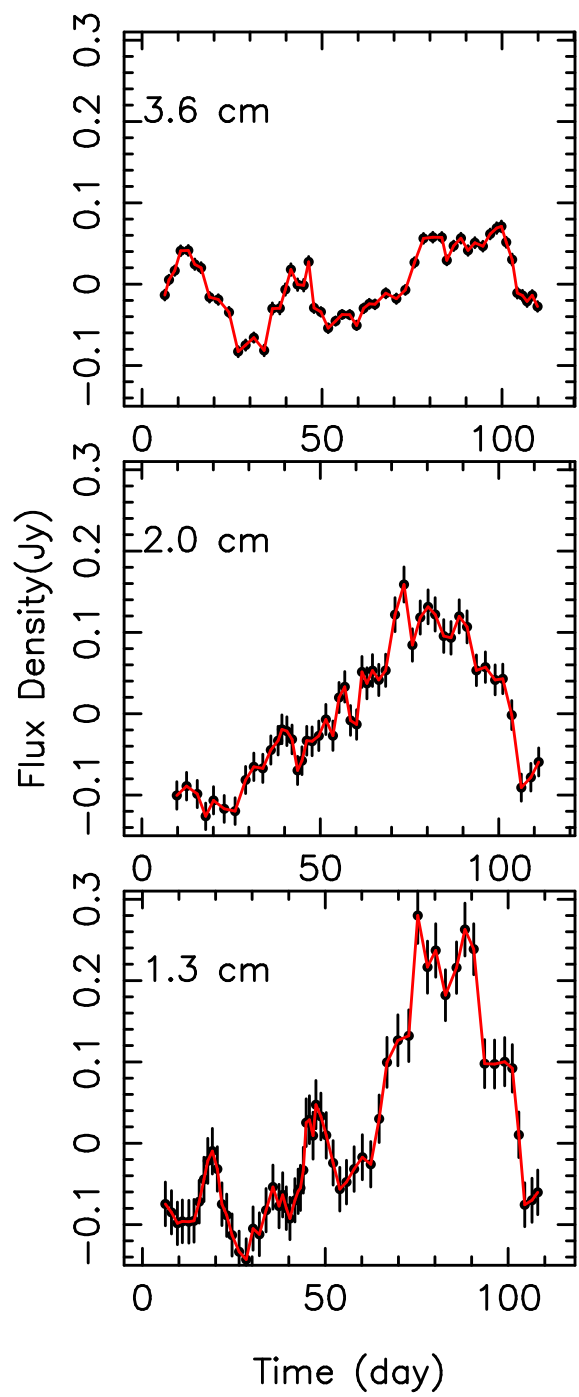
\end{document}